\documentclass[11pt,a4paper]{article}
\pdfoutput=1
\usepackage{jheppub}
\usepackage{tikz}
\usetikzlibrary{intersections, calc, arrows.meta, bending}
\usepackage{subcaption}
\usepackage{simpler-wick}

\DeclareMathOperator{\tr}{tr}

\newcommand{\ri}{\mathrm{i}}
\renewcommand{\th}{\theta}

\newcommand{\cob}{\delta}

\newcommand{\vep}{\varepsilon}

\newcommand{\hf}{\frac{1}{2}}

\newcommand{\del}{\partial}

\newcommand{\bra}{\langle}
\newcommand{\ket}{\rangle}
\newcommand{\la}{\lambda}

\newcommand{\bt}{\beta}
\newcommand{\ga}{\gamma}

\newcommand{\rt}[1]{\sqrt{#1}}
\newcommand{\cO}{\mathcal{O}}

\newcommand{\cN}{\mathcal{N}}
\newcommand{\cM}{\mathcal{M}}

\newcommand{\cC}{\mathcal{C}}

\newcommand{\cJ}{\mathcal{J}}
\newcommand{\mZ}{\mathbb{Z}}

\newcommand{\bJT}{\beta_\text{JT}}

\begin{document}
\title{Non-perturbative corrections in the semi-classical limit of 
double-scaled SYK}

\author{Kazumi Okuyama}

\affiliation{Department of Physics, 
Shinshu University, 3-1-1 Asahi, Matsumoto 390-8621, Japan}

\emailAdd{kazumi@azusa.shinshu-u.ac.jp}

\abstract{We study the disk partition function of
double-scaled SYK model (DSSYK)
in the small $\lambda$ limit,
where $\lambda=-\log q$ is the coupling of DSSYK.
We find that the partition function receives non-perturbative corrections 
in $\lambda$, which can be resummed 
by the cubic power of the Dedekind eta function
in a certain low temperature limit.
We also discuss a possible bulk interpretation of our findings.}

\maketitle

\section{Introduction}

The double-scaled SYK model (DSSYK)
is a useful toy model of the holographic duality between
a quantum mechanical system and a bulk quantum gravity.
Since DSSYK is exactly solvable by the technique of
the chord diagrams and the $q$-deformed oscillators
\cite{Berkooz:2018jqr},
we can learn many features of the holographic duality
in great detail.
For instance, the Fock space of the $q$-deformed oscillators,
known as the chord Hilbert space,
can be identified as the Hilbert space of
bulk quantum gravity \cite{Lin:2022rbf},
and the bulk geodesic lengths become discretized in units of
$\la=-\log q$ \cite{Lin:2022rbf,Jafferis:2022wez,Okuyama:2023byh}.

There are several proposals for the bulk dual of DSSYK.
In \cite{Narovlansky:2023lfz,Verlinde:2024znh,Verlinde:2024zrh,Susskind:2021esx,Susskind:2022dfz,Susskind:2022bia,Susskind:2023hnj,Rahman:2023pgt,Rahman:2024vyg,Rahman:2024iiu,Sekino:2025bsc},
it is argued that DSSYK is dual to de Sitter space 
in a certain limit.
In \cite{Blommaert:2024ydx,Blommaert:2024whf},
the bulk dual of DSSYK is identified as the sine dilaton
gravity and this proposal is checked at the one-loop
level \cite{Bossi:2024ffa}.
However, this identification is questioned in 
\cite{Collier:2025pbm}
since the sine dilaton gravity is
dual to the complex Liouville string, not DSSYK, according to
\cite{Collier:2024kmo}.
It is fair to say that the bulk dual of DSSYK is not 
completely understood yet.

In this paper, we consider the disk partition function
$Z(\bt)$ of DSSYK in the semi-classical, $\la\to0$ limit.
We will not assume any particular bulk dual of DSSYK.
Rather, we will try to extract the bulk picture of DSSYK
from the honest calculation of the semi-classical limit of 
$Z(\bt)$.
It is known that the disk partition function
of DSSYK reduces to that of JT gravity
in the semi-classical limit at low energy \cite{Berkooz:2018jqr}. 
We find that $Z(\bt)$ of DSSYK contains
non-perturbative corrections on top of the JT gravity result,
and such corrections can be nicely resummed 
into the cubic power of the Dedekind eta function.
In in section \ref{sec:discussion},
we will discuss a possible bulk interpretation of our result.

This paper is organized as follows.
In section \ref{sec:disk}, we review the 
result \eqref{eq:exact-Z} of the
disk partition function $Z(\bt)$ of DSSYK
and rewrite the measure $\mu(\th)$ as a sum of Gaussians \eqref{eq:gauss-sum}.
In section \ref{sec:limit}, we consider various limits of
$Z(\bt)$: (i) naive semi-classical limit (\S\ref{sec:naive}),
(ii) low energy limit (\S\ref{sec:lowE}), and (iii)
low temperature limit (\S\ref{sec:lowT}).
We find that the non-perturbative corrections can be resummed 
in terms of the Dedekind eta function as
\eqref{eq:Z-eta1} in the low energy limit
and \eqref{eq:Z-eta2} in the low temperature limit.
Finally, in section \ref{sec:discussion} we discuss
a possible bulk interpretation of our findings.

\section{Disk partition function of DSSYK}\label{sec:disk}
In this section, we review the exact result of the disk partition function
of DSSYK \cite{Berkooz:2018jqr}.
The SYK model is defined by the $p$-body Hamiltonian for 
the $N$ Majorana fermions $\psi_i~(i=1,\cdots,N)$
\cite{Sachdev1993,Kitaev1,Kitaev2}
\begin{equation}
\begin{aligned}
 H=\ri^{p/2}\sum_{1\leq i_1<\cdots<i_p\leq N}J_{i_1\cdots i_p}\psi_{i_1}\cdots
\psi_{i_p},
\end{aligned} 
\end{equation}
with the random coupling $J_{i_1\cdots i_p}$.
This coupling is assumed to be Gaussian random with zero mean 
and the variance is given by
\begin{equation}
\begin{aligned}
 \Bigl\bra J_{i_1\cdots i_p}^2\Bigr\ket=\frac{\cJ^2}{\la \binom{N}{p}}.
\end{aligned} 
\end{equation}
DSSYK is defined by taking the limit
\begin{equation}
\begin{aligned}
 p,N\to\infty \quad \text{with}\quad \la=\frac{2p^2}{N}:~\text{fixed}.
\end{aligned} 
\label{eq:la-def}
\end{equation}
In this scaling limit, the computation of the moment $\bra\tr H^k\ket$
boils down to the counting problem of the intersection numbers of the
chord diagrams.
As shown in \cite{Berkooz:2018jqr},
this counting problem can be exactly solved by introducing the
transfer matrix $T=a+a^\dagger$
acting on the chord Hilbert space, where $a,a^\dagger$
are the $q$-deformed oscillators with $q=e^{-\la}$.
Using this technique,
we can write down the exact form of the disk partition function of DSSYK
\begin{equation}
\begin{aligned}
 Z(\bt)&=\int_0^\pi\frac{d\th}{2\pi}\mu(\th)e^{-\bt E(\th)},
\end{aligned} 
\label{eq:exact-Z}
\end{equation}
where $E(\th)$ is the eigenvalue of the transfer matrix $T$
\begin{equation}
\begin{aligned}
 E(\th)=-E_0\cos\th,\quad E_0=\frac{2\cJ}{\rt{\la(1-q)}},
\end{aligned} 
\label{eq:E-th}
\end{equation}
and the measure factor $\mu(\th)$ is given by
\footnote{We use the standard notation for the $q$-Pochhammer symbol
\begin{equation}
\begin{aligned}
 (a;q)_\infty=\prod_{k=0}^\infty (1-aq^k),\quad
(a_1,\cdots,a_s;q)_\infty=\prod_{i=1}^s(a_i;q)_\infty,\quad
(e^{\pm2\ri\th};q)_\infty=
(e^{2\ri\th},e^{-2\ri\th};q)_\infty.
\end{aligned} 
\end{equation}}
\begin{equation}
\begin{aligned}
 \mu(\th)=(q,e^{\pm2\ri\th};q)_\infty.
\end{aligned} 
\label{eq:mu-th}
\end{equation}

Let us summarize some useful properties of the spectrum $E(\th)$
and the measure $\mu(\th)$
appearing in the disk partition function \eqref{eq:exact-Z}.
First of all, $E(\th)$ is bounded from below and above
\begin{equation}
\begin{aligned}
 -E_0\leq E(\th)\leq E_0,
\end{aligned} 
\end{equation}
and the lower end $E=-E_0$ and the upper end $E=E_0$ correspond to $\th=0$ 
and $\th=\pi$, respectively.
One can see that $E(\th)$
in \eqref{eq:E-th} and $\mu(\th)$
in \eqref{eq:mu-th} have the following symmetries
\begin{equation}
\begin{aligned}
 E(\th)&=E(-\th)=-E(\th+\pi),\\
\mu(\th)&=\mu(-\th)=\mu(\th+\pi),
\end{aligned} 
\label{eq:period}
\end{equation}
which implies that $Z(\bt)$ in \eqref{eq:exact-Z}
is an even function of $\bt$
\begin{equation}
\begin{aligned}
 Z(-\bt)=Z(\bt).
\end{aligned} 
\label{eq:even}
\end{equation}
One can also see that $\mu(\th)$ is a positive definite measure
on the real $\th$-axis
\begin{equation}
\begin{aligned}
 \mu(\th)\geq0,\quad (\th\in\mathbb{R}),
\end{aligned} 
\label{eq:positive-mu}
\end{equation}
where we extended the support of $\mu(\th)$ beyond $0<\th<\pi$
using the periodicity of $\mu(\th)$ in \eqref{eq:period}.

In this paper, we consider the semi-classical, small $\la$ limit of the disk
partition function $Z(\bt)$ in \eqref{eq:exact-Z}.
To study this limit, it is useful to rewrite $\mu(\th)$
in terms of the Jacobi theta function
\begin{equation}
\begin{aligned}
 \mu(\th)=2q^{-\frac{1}{8}}\sin\th\, \vartheta_1
\Bigl(\frac{\th}{\pi},\frac{\ri\la}{2\pi}\Bigr),
\end{aligned} 
\end{equation}
where $\vartheta_1(v,\tau)$ is given by
\begin{equation}
\begin{aligned}
 \vartheta_1(v,\tau)&=2q^{\frac{1}{8}}\sin(\pi v)
\prod_{n=1}^\infty (1-q^n)(1-e^{2\pi\ri v}q^n)
(1-e^{-2\pi\ri v}q^n)\\
&=-\ri\sum_{n\in\mZ}(-1)^n q^{\hf(n+\hf)^2}e^{\pi\ri v(2n+1)},
\end{aligned} 
\label{eq:sum-theta}
\end{equation}
with $q=e^{2\pi\ri\tau}$.
The small $\la$ limit of $\mu(\th)$
can be analyzed by the S-transformation of the Jacobi theta function
\begin{equation}
\begin{aligned}
 \vartheta_1(v,\tau)=\ri (-\ri\tau)^{-\hf}
e^{-\frac{\pi\ri v^2}{\tau}}\vartheta_1\Bigl(\frac{v}{\tau},-\frac{1}{\tau}\Bigr).
\end{aligned} 
\end{equation}
Alternatively, we can apply the Poisson resummation 
formula to the second line of
\eqref{eq:sum-theta}.
\begin{equation}
\begin{aligned}
 \sum_{n\in\mZ}f(n)=\sum_{j\in\mZ}\int_{-\infty}^\infty dx e^{2\pi\ri jx}f(x).
\end{aligned} 
\end{equation}
In this way, we find that $\mu(\th)$ is written as a sum of Gaussian factors
\begin{equation}
\begin{aligned}
 \mu(\th)=\cC\sin\th\sum_{j\in\mZ}(-1)^je^{-\frac{2}{\la}(\th-\th_j)^2},
\end{aligned} 
\label{eq:gauss-sum}
\end{equation}
where $\cC$ and $\th_j$ are given by
\begin{equation}
\begin{aligned}
 \cC=2 q^{-\frac{1}{8}}\rt{\frac{2\pi}{\la}},\qquad \th_j=\pi\Bigl(j+\hf\Bigr).
\end{aligned} 
\label{eq:C-thj}
\end{equation}
This expression \eqref{eq:gauss-sum}
of $\mu(\th)$ as a sum of Gaussian factors
has been already appeared in \cite{Verlinde:2024znh}
and a possible bulk interpretation of this expression was discussed
in \cite{Verlinde:2024znh,Blommaert:2024ydx,Blommaert:2024whf}:
it is argued that the Gaussian factors in \eqref{eq:gauss-sum}
come from the conical defects in the bulk, but the 
bulk interpretation of the sign $(-1)^j$ 
is not clearly understood.

From the boundary DSSYK viewpoint, the necessity
of this sign $(-1)^j$ is clear from the positivity of $\mu(\th)$
in \eqref{eq:positive-mu}.
The sign $(-1)^j$ in \eqref{eq:gauss-sum} is necessary to cancel the extra sign
coming from 
$\sin\th$ at $\th=\th_j$
\begin{equation}
\begin{aligned}
 \sin\th_j=(-1)^j.
\end{aligned} 
\end{equation}
More generally, in the range of $\th\in[\th_j-\pi/2,\th_j+\pi/2]$
of length $\pi$ around $\th=\th_j$, $\mu(\th)$ is approximated by the
$j^\text{th}$ term of \eqref{eq:gauss-sum}
in the small $\la$ limit
\begin{equation}
\begin{aligned}
 \mu(\th)&\approx \cC\sin\th\,(-1)^j e^{-\frac{2}{\la}(\th-\th_j)^2}
=\cC\cos(\th-\th_j)e^{-\frac{2}{\la}(\th-\th_j)^2}.
\end{aligned} 
\end{equation}
Indeed, this is positive in the range $\th\in[\th_j-\pi/2,\th_j+\pi/2]$
in question.

In the rest of this paper, we study various limits of $Z(\bt)$
using the expression of $\mu(\th)$ in \eqref{eq:gauss-sum}.

\section{Various limits of disk partition function}\label{sec:limit}

\subsection{Semi-classical limit with $\cJ\sim\cO(\la^0)$}\label{sec:naive}
First, let us consider 
the naive semi-classical limit \cite{Maldacena:2016hyu,Goel:2023svz}
\begin{equation}
\begin{aligned}
 \la\to0,\quad \cJ\sim \cO(\la^0).
\end{aligned} 
\end{equation}
Since $E_0$ in \eqref{eq:E-th} is order $\cO(\la^{-1})$
in this limit, it is natural to set
\begin{equation}
\begin{aligned}
 \bt E_0=\frac{2\bt_s}{\la}\quad\text{with}\quad \bt_s:~\text{fixed}.
\end{aligned} 
\end{equation}
Then, $Z(\bt)$ in \eqref{eq:exact-Z} becomes
\begin{equation}
\begin{aligned}
 Z(\bt)&=\int_0^\pi\frac{d\th}{2\pi}\cC\sin\th\sum_{j\in\mZ}(-1)^j
e^{-\frac{2}{\la}(\th-\th_j)^2+\frac{2\bt_s}{\la}\cos\th}\\
&\approx \int_0^\pi\frac{d\th}{2\pi}\cC\sin\th
e^{-\frac{2}{\la}(\th-\frac{\pi}{2})^2+\frac{2\bt_s}{\la}\cos\th}.
\end{aligned} 
\end{equation}
When going to the second line, we took only the $j=0$ term
which is dominant in the small $\la$ limit.
This integral can be evaluated by the saddle point method,
and the saddle point $\th=\th_*$ is given by
\begin{equation}
\begin{aligned}
 \bt_s=\frac{\pi-2\th_*}{\sin\th_*}.
\end{aligned} 
\end{equation}
This reproduces the result of \cite{Maldacena:2016hyu,Goel:2023svz}, 
as expected.

\subsection{Low energy limit with $\cJ\sim\cO(\la^{-1})$}\label{sec:lowE}
As discussed in \cite{Berkooz:2018jqr},
DSSYK reduces to JT gravity by zooming in
on the edge $E=-E_0$ of the spectrum in the semi-classical limit
\begin{equation}
\begin{aligned}
 \la,\th\to0\quad\text{with}\quad k=\frac{\th}{\la}:~\text{fixed}.
\end{aligned} 
\label{eq:JT-lim}
\end{equation}
We also scale $\cJ$ as $\cO(\la^{-1})$ and set
\begin{equation}
\begin{aligned}
 \bt E_0=\frac{\bJT}{\la^2}\quad\text{with}\quad\bJT:~\text{fixed}.
\end{aligned} 
\end{equation}
In this limit, the Boltzmann factor becomes
\begin{equation}
\begin{aligned}
 e^{-\bt E(\th)}=e^{\bt E_0\cos(\la k)}\approx e^{\bt E_0-\hf\bJT k^2}.
\end{aligned} 
\label{eq:dispersion}
\end{equation}
We rewrite $\mu(\th)$ in \eqref{eq:gauss-sum}
by combining the terms at $j$ and $-j-1$.
Using $\th_{-j-1}=-\th_j$, we find
\begin{equation}
\begin{aligned}
 &(-1)^j e^{-\frac{2}{\la}(\th-\th_j)^2}+(-1)^{-j-1}e^{-\frac{2}{\la}(\th-\th_{-j-1})^2}
=(-1)^j e^{-\frac{2}{\la}(\th^2+\th_j^2)}
2\sinh\left(\frac{4\th\th_j}{\la}\right),
\end{aligned} 
\end{equation}
and $\mu(\th)$ is written as
\begin{equation}
\begin{aligned}
 \mu(\th)=\cC \sin\th e^{-\frac{2\th^2}{\la}}\sum_{j=0}^\infty (-1)^j
e^{-\frac{2\th_j^2}{\la}}2\sinh\left(\frac{4\th\th_j}{\la}\right).
\end{aligned} 
\end{equation}
In the limit \eqref{eq:JT-lim}, this reduces to
\begin{equation}
\begin{aligned}
 \mu(\th)= \la \cC 
\sum_{j=0}^\infty (-1)^j
\tilde{q}^{\hf(j+\hf)^2}2k\sinh\bigl[2\pi k(2j+1)\bigr],
\end{aligned} 
\label{eq:mu-low}
\end{equation}
where $\tilde{q}$ is the S-transform of $q=e^{-\la}$
\begin{equation}
\begin{aligned}
 \tilde{q}=e^{-\frac{4\pi^2}{\la}}.
\end{aligned} 
\end{equation}
The $j=0$ term of \eqref{eq:mu-low}
correctly reproduces 
the spectral density of 
the Schwarzian theory
\cite{Stanford:2017thb}
\begin{equation}
\begin{aligned}
 \mu(\th)\sim 2k\sinh(2\pi k).
\end{aligned} 
\end{equation}
Together with the dispersion relation $E(k)=\hf k^2$ 
in \eqref{eq:dispersion},
this reproduces the partition function of JT gravity
\cite{Saad:2019lba}
\begin{equation}
\begin{aligned}
 \int_0^\infty \frac{dk}{2\pi}
2k\sinh(2\pi k)e^{-\hf\bJT k^2}=2\pi
Z_\text{JT}(\bJT),
\end{aligned} 
\end{equation}
where $Z_\text{JT}(\bJT)$ is given by
\begin{equation}
\begin{aligned}
 Z_\text{JT}(\bJT)=\frac{e^{\frac{2\pi^2}{\bJT}}}{\rt{2\pi\bJT^3}}.
\end{aligned} 
\label{eq:ZJT}
\end{equation}

The $j\ne0$ terms of \eqref{eq:mu-low}
can be thought of as the non-perturbative corrections to the JT gravity
result \eqref{eq:ZJT}.
Plugging $E(\th)$ in \eqref{eq:dispersion} and $\mu(\th)$
in \eqref{eq:mu-low} into the definition of $Z(\bt)$
in \eqref{eq:exact-Z}, we find
\begin{equation}
\begin{aligned}
 Z(\bt)&= \la\cC e^{\bt E_0}\int_0^\infty \frac{\la dk}{2\pi} 
\sum_{j=0}^\infty (-1)^j
\tilde{q}^{\hf(j+\hf)^2}2k\sinh\bigl[2\pi k(2j+1)\bigr] e^{-\hf\bJT k^2}\\
&=2\pi \la^2\cC e^{\bt E_0} 
\sum_{j=0}^\infty (-1)^j
\tilde{q}^{\hf(j+\hf)^2} (2j+1)\frac{e^{\frac{2\pi^2(2j+1)^2}{\bt_\text{JT}}}}{\rt{2\pi \bt_{\text{JT}}^3}}.
\end{aligned} 
\label{eq:ZJT-expand}
\end{equation}
It turns out that this sum over $j$ can be performed in a closed form,
thanks to the formula\footnote{This identity can be easily obtained 
by taking the $v\to0$ limit of $\frac{\vartheta_1(v,\tau)}{2\pi v}$ 
in \eqref{eq:sum-theta}.}
\begin{equation}
\begin{aligned}
 \sum_{j=0}^\infty (-1)^j q^{\hf(j+\hf)^2}(2j+1)=
\eta(q)^3,
\end{aligned} 
\label{eq:sum-eta}
\end{equation}
where $\eta(q)$ denotes the Dedekind eta function
\begin{equation}
\begin{aligned}
 \eta(q)=q^{\frac{1}{24}}\prod_{n=1}^\infty (1-q^n).
\end{aligned} 
\end{equation}
Finally we arrive at the closed form of the partition function
in the low energy limit \eqref{eq:JT-lim}
\begin{equation}
\begin{aligned}
 Z(\bt)=
2\pi \la^2\cC \frac{e^{\bt E_0}}{\rt{2\pi\bt_{\text{JT}}^3}}\eta
\Bigl(\tilde{q}e^{\frac{16\pi^2}{\bt_{\text{JT}}}}\Bigr)^3 .
\end{aligned} 
\label{eq:Z-eta1}
\end{equation}
We should stress that this expression contains
non-perturbative corrections to the JT gravity result 
\eqref{eq:ZJT}. 
Indeed, the leading term of the small $\tilde{q}$ expansion of $\eta
\Bigl(\tilde{q}e^{\frac{16\pi^2}{\bt_{\text{JT}}}}\Bigr)^3$
reproduces the JT gravity partition function \eqref{eq:ZJT}
\begin{equation}
\begin{aligned}
 \Bigl(\tilde{q}e^{\frac{16\pi^2}{\bt_{\text{JT}}}}\Bigr)^{\frac{3}{24}}
=\tilde{q}^{\frac{1}{8}}e^{\frac{2\pi^2}{\bJT}}.
\end{aligned} 
\end{equation}
The higher order terms in the small
$\tilde{q}$ expansion of \eqref{eq:Z-eta1}
give rise to the corrections
of order $\cO(\tilde{q})=\cO(e^{-\frac{4\pi^2}{\la}})$,
which are non-perturbative in $\la$.

\subsection{Low temperature limit}\label{sec:lowT}
As discussed in \cite{Berkooz:2018jqr},
the $\th$-integral of the 
exact partition function \eqref{eq:exact-Z}
can be evaluated without approximation
\begin{equation}
\begin{aligned}
 Z(\bt)=\sum_{r=0}^\infty (-1)^r q^{\hf r(r+1)}(2r+1)\frac{2I_{2r+1}(\bt E_0)}{\bt E_0},
\end{aligned} 
\label{eq:Z-Bessel}
\end{equation}
where $I_\nu(z)$ denotes the modified Bessel function of the first kind.
In this subsection, we consider the 
low temperature limit of the exact result \eqref{eq:Z-Bessel}
\begin{equation}
\begin{aligned}
 \bt E_0\gg1.
\end{aligned}
\label{eq:lowT} 
\end{equation}
It is well-known that the modified Bessel function $I_\nu(z)$
grows exponentially at large $z$
\begin{equation}
\begin{aligned}
 I_\nu(z)=\frac{e^z}{\rt{2\pi z}}\Bigl[1+\cO(z^{-1})\Bigr],\quad (z\gg1).
\end{aligned} 
\end{equation}
However, this is not the end of the story.
$I_\nu(z)$ contains exponentially small corrections
as well \footnote{See e.g. \url{https://dlmf.nist.gov/10.40}.}
\begin{equation}
\begin{aligned}
 I_\nu(z)=\frac{e^z}{\rt{2\pi z}}\Bigl[1+\cO(z^{-1})\Bigr]
+e^{\pi\ri\nu}\frac{e^{-z}}{\rt{-2\pi z}}\Bigl[1+\cO(z^{-1})\Bigr],\quad (z\gg1).
\end{aligned} 
\label{eq:I-asymp}
\end{equation}
From this behavior, 
we find the low temperature limit \eqref{eq:lowT} of 
the modified Bessel function $I_{2r+1}(\bt E_0)$ in 
\eqref{eq:Z-Bessel}
\begin{equation}
\begin{aligned}
 I_{2r+1}(\bt E_0)\approx\frac{e^{\bt E_0}}{\rt{2\pi \bt E_0}}-
\frac{e^{-\bt E_0}}{\rt{-2\pi \bt E_0}}.
\end{aligned} 
\end{equation} 
Then, using \eqref{eq:sum-eta} the partition function becomes
\begin{equation}
\begin{aligned}
 Z(\bt)&\approx 2q^{-\frac{1}{8}}\sum_{r=0}^\infty (-1)^r q^{\hf(r+\hf)^2}(2r+1)
\left[\frac{e^{\bt E_0}}{\rt{2\pi(\bt E_0)^3}}+
\frac{e^{-\bt E_0}}{\rt{2\pi(-\bt E_0)^3}}\right]\\
&=2q^{-\frac{1}{8}}\eta(q)^3
\left[\frac{e^{\bt E_0}}{\rt{2\pi(\bt E_0)^3}}+
\frac{e^{-\bt E_0}}{\rt{2\pi(-\bt E_0)^3}}\right].
\end{aligned} 
\label{eq:Zlow-eta}
\end{equation}
Finally, using the modular transformation of the eta function
\begin{equation}
\begin{aligned}
 \eta(q)=\rt{\frac{2\pi}{\la}}\eta(\tilde{q}),
\end{aligned} 
\end{equation}
we can rewrite \eqref{eq:Zlow-eta} as
\begin{equation}
\begin{aligned}
 Z(\bt)= 2q^{-\frac{1}{8}}\Bigl(\frac{2\pi}{\la}\Bigr)^{\frac{3}{2}}
\eta(\tilde{q})^3 
\left[\frac{e^{\bt E_0}}{\rt{2\pi(\bt E_0)^3}}+
\frac{e^{-\bt E_0}}{\rt{2\pi(-\bt E_0)^3}}\right].
\end{aligned} 
\label{eq:Z-eta2}
\end{equation}
One can check that the large $\bJT$
limit of \eqref{eq:Z-eta1} agrees with the first term of 
\eqref{eq:Z-eta2}.
It would be possible to reproduce \eqref{eq:Z-eta1}
by performing a resummation of the correction terms of $I_\nu(z)$,
indicated by $\cO(z^{-1})$ in \eqref{eq:I-asymp}.
We leave this as an interesting future problem.

The second term of \eqref{eq:Z-eta2}
comes from the upper end of the spectrum $E=E_0$,
which corresponds to $\th=\pi$.
This second contribution is required from
the symmetry \eqref{eq:even}.
Note that this term comes with a factor of $(-1)^{-\frac{3}{2}}=\ri$,
which suggests that $\th=\pi$ is an unstable saddle point.\footnote{
Note that the expectation value of 
BPS Wilson loops in $\cN=4$ super Yang-Mills is also given by the modified Bessel
function \cite{Erickson:2000af}.
In this context, the second term of \eqref{eq:I-asymp}
is interpreted as a contribution of unstable worldsheet
instanton in the bulk $AdS_5\times S^5$ \cite{Drukker:2006ga}.
}
In fact, expanding the Boltzmann factor around $\th=\pi$, we find a 
wrong sign Gaussian
\begin{equation}
\begin{aligned}
 e^{-\bt E(\pi-\phi)}=e^{-\bt E_0\cos\phi}
=e^{-\bt E_0+\hf\bt E_0\phi^2+\cO(\phi^4)},
\end{aligned} 
\end{equation}
where we have set $\th=\pi-\phi~(|\phi|\ll1)$.
In order to make the $\phi$-integral convergent, we have to rotate the contour
of $\phi$ to the imaginary direction
\begin{equation}
\begin{aligned}
 \phi=\ri x,\quad (x\in\mathbb{R}).
\end{aligned} 
\end{equation}
This explains the appearance of the factor of 
$(-1)^{-\frac{3}{2}}$ in the second term of \eqref{eq:Z-eta2}.

\section{Discussion}\label{sec:discussion}
In this paper, we have studied the semi-classical $\la\to0$ limit
of the disk partition function of DSSYK.
We found that the non-perturbative corrections in $\la$
can be resummed by the cubic power of the Dedekind eta function 
in the low energy limit \eqref{eq:Z-eta1}
and the low temperature limit \eqref{eq:Z-eta2}.
In this concluding section, we discuss a possible bulk interpretation
of our result \eqref{eq:Z-eta1}. 

We have seen in \eqref{eq:ZJT} that
in the low energy, small $\la$ limit,
the partition function of DSSYK reduces to that of JT gravity with negative
cosmological constant. 
As discussed in \cite{Saad:2019lba},
the partition function
of JT gravity 
depends on the boundary value
of the dilaton $\Phi$
\begin{equation}
\begin{aligned}
 \Phi\big|_{\del \cM}=\frac{\ga}{\vep}\quad(\vep\to0),
\end{aligned} 
\end{equation}
where $\cM$ is the bulk spacetime and $\del\cM$ is the 
boundary of the asymptotic $AdS_2$ part of $\cM$
at $z=\vep$ in the Poincar\'{e} coordinate.
Note that $\ga$ plays the role of the coupling of Schwarzian mode.
Including the $\ga$-dependence, the disk partition function of
JT gravity becomes 
\begin{equation}
\begin{aligned}
 Z_{\text{JT}}(\bt_{\text{JT}})\propto\frac{1}{\rt{2\pi\bt_\text{JT}^3}}
e^{\frac{2\pi^2\ga}{\bt_{\text{JT}}}}.
\end{aligned} 
\end{equation}
Here the proportionality constant is regularization dependent \cite{Stanford:2017thb,Saad:2019lba}.
For our purpose, it is convenient choose this constant as
\begin{equation}
\begin{aligned}
 Z_{\text{JT}}(\bt_{\text{JT}},\ga)=
\frac{\rt{\ga}}{\rt{2\pi\bt_\text{JT}^3}}e^{\frac{2\pi^2\ga}{\bt_{\text{JT}}}}.
\end{aligned} 
\end{equation}
Then, our result \eqref{eq:Z-eta1} of the low energy limit of 
$Z(\bt)$ is written as
\begin{equation}
\begin{aligned}
Z(\bt)=2\pi\la^2\cC e^{\bt E_0} \sum_{j=0}^\infty (-1)^j \tilde{q}^{\frac{\ga_j}{8}}
Z_{\text{JT}}(\bt_{\text{JT}},\ga_j),
\end{aligned} 
\label{eq:Z-gaj}
\end{equation}
with $\ga_j$ being
\begin{equation}
\begin{aligned}
 \ga_j=(2j+1)^2.
\end{aligned} 
\end{equation}
This expression \eqref{eq:Z-gaj}
suggests that DSSYK is holographically dual to a
superposition of the boundary conditions of the dilaton
in JT gravity with negative cosmological constant,
at least at low energy in the semi-classical limit.
This is formally realized by inserting the $\cob$-function
\begin{equation}
\begin{aligned}
 \sum_{j=0}^\infty (-1)^j\tilde{q}^{\frac{\ga_j}{8}}
\cob\Bigl(\Phi|_{\del\cM}-\frac{\ga_j}{\vep}\Bigr)
\end{aligned} 
\end{equation}
into the JT gravity path integral.
It is not clear to us how this picture is related to the sine dilaton gravity.
In \cite{Berkooz:2024ifu}, the low temperature limit of DSSYK is
discussed from the viewpoint of Schwarzian theory.
It would be interesting to study the relation between
the result in \cite{Berkooz:2024ifu} and our \eqref{eq:Z-gaj}.

We can improve \eqref{eq:Z-gaj} by including the contribution
from $\th=\pi$. As we have seen in \eqref{eq:Z-eta2},
the contribution
from $\th=\pi$ is the analytic continuation $\bt\to-\bt$
of the contribution from $\th=0$.
Thus the improved version of \eqref{eq:Z-gaj} is
\begin{equation}
\begin{aligned}
 Z(\bt)=2\pi\la^2\cC\sum_{j=0}^\infty (-1)^j\tilde{q}^{\frac{\ga_j}{8}}
\Bigl[e^{\bt E_0}Z_{\text{JT}}(\bt_{\text{JT}},\ga_j)
+e^{-\bt E_0}Z_{\text{JT}}(-\bt_{\text{JT}},\ga_j)\Bigr].
\end{aligned} 
\label{eq:Z-sym}
\end{equation}
It would be interesting to understand the bulk interpretation of the second term coming
from $\th=\pi$.

In this paper, we have only considered the low energy and/or low temperature
limit of DSSYK.
It would be interesting to study the high temperature limit of DSSYK, in
view of its possible relation to de Sitter space.
In \cite{Okuyama:2023iwu}, it was found that the small $\bt$
expansion of the free energy $\log Z(\bt)$ of DSSYK
has a finite radius of convergence, which is related to the zero
of $Z(\bt)$ on the imaginary $\bt$-axis.
It would be interesting to understand the analytic structure of 
$Z(\bt)$ on the complex $\bt$-plane and its relation to
de Sitter space, if any.

Finally, we would like to comment on a possible interpretation of 
\eqref{eq:Z-Bessel} and \eqref{eq:Z-sym} in the sine dilaton gravity.\footnote{
We would like to thank the anonymous referee of JHEP for suggesting 
this interpretation.}
As discussed in \cite{Blommaert:2025avl}, the modified Bessel functions
in \eqref{eq:Z-Bessel} naturally appear in the closed-channel canonical quantization of sine dilaton gravity
and they are interpreted as the physical trumpets 
which satisfy the radial Wheeler--DeWitt equation
of the theory. 
Also, it is argued in \cite{Blommaert:2024whf}
that the second term of \eqref{eq:Z-sym}, corresponding to the expansion
around $\Phi=\pi$ in the sine dilaton gravity, is related to
the de Sitter JT gravity regime of sine dilaton gravity.
It would be interesting to understand this relation better.

\acknowledgments
This work was supported
in part by JSPS Grant-in-Aid for Transformative Research Areas (A) 
``Extreme Universe'' 21H05187 and JSPS KAKENHI 22K03594.
\bibliography{paper}

\providecommand{\href}[2]{#2}\begingroup\raggedright\begin{thebibliography}{10}

\bibitem{Berkooz:2018jqr}
M.~Berkooz, M.~Isachenkov, V.~Narovlansky, and G.~Torrents, ``{Towards a full
  solution of the large N double-scaled SYK model},''
  \href{https://doi.org/10.1007/JHEP03(2019)079}{{ JHEP} {\bfseries 03} (2019)
  079}, \href{https://arxiv.org/abs/1811.02584}{{\ttfamily arXiv:1811.02584
  [hep-th]}}.

\bibitem{Lin:2022rbf}
H.~W. Lin, ``{The bulk Hilbert space of double scaled SYK},''
  \href{https://doi.org/10.1007/JHEP11(2022)060}{{ JHEP} {\bfseries 11} (2022)
  060}, \href{https://arxiv.org/abs/2208.07032}{{\ttfamily arXiv:2208.07032
  [hep-th]}}.

\bibitem{Jafferis:2022wez}
D.~L. Jafferis, D.~K. Kolchmeyer, B.~Mukhametzhanov, and J.~Sonner,
  ``{Jackiw-Teitelboim gravity with matter, generalized eigenstate
  thermalization hypothesis, and random matrices},''
  \href{https://doi.org/10.1103/PhysRevD.108.066015}{{ Phys. Rev. D} {\bfseries
  108} no.~6, (2023) 066015},
  \href{https://arxiv.org/abs/2209.02131}{{\ttfamily arXiv:2209.02131
  [hep-th]}}.

\bibitem{Okuyama:2023byh}
K.~Okuyama, ``{End of the world brane in double scaled SYK},''
  \href{https://doi.org/10.1007/JHEP08(2023)053}{{ JHEP} {\bfseries 08} (2023)
  053}, \href{https://arxiv.org/abs/2305.12674}{{\ttfamily arXiv:2305.12674
  [hep-th]}}.

\bibitem{Narovlansky:2023lfz}
V.~Narovlansky and H.~Verlinde, ``{Double-scaled SYK and de Sitter
  Holography},'' \href{https://arxiv.org/abs/2310.16994}{{\ttfamily
  arXiv:2310.16994 [hep-th]}}.

\bibitem{Verlinde:2024znh}
H.~Verlinde, ``{Double-scaled SYK, Chords and de Sitter Gravity},''
  \href{https://arxiv.org/abs/2402.00635}{{\ttfamily arXiv:2402.00635
  [hep-th]}}.

\bibitem{Verlinde:2024zrh}
H.~Verlinde and M.~Zhang, ``{SYK Correlators from 2D Liouville-de Sitter
  Gravity},'' \href{https://arxiv.org/abs/2402.02584}{{\ttfamily
  arXiv:2402.02584 [hep-th]}}.

\bibitem{Susskind:2021esx}
L.~Susskind, ``{Entanglement and Chaos in De Sitter Space Holography: An SYK
  Example},'' \href{https://doi.org/10.22128/jhap.2021.455.1005}{{ JHAP}
  {\bfseries 1} no.~1, (2021) 1--22},
  \href{https://arxiv.org/abs/2109.14104}{{\ttfamily arXiv:2109.14104
  [hep-th]}}.

\bibitem{Susskind:2022dfz}
L.~Susskind, ``{Scrambling in Double-Scaled SYK and De Sitter Space},''
  \href{https://arxiv.org/abs/2205.00315}{{\ttfamily arXiv:2205.00315
  [hep-th]}}.

\bibitem{Susskind:2022bia}
L.~Susskind, ``{De Sitter Space, Double-Scaled SYK, and the Separation of
  Scales in the Semiclassical Limit},''
  \href{https://arxiv.org/abs/2209.09999}{{\ttfamily arXiv:2209.09999
  [hep-th]}}.

\bibitem{Susskind:2023hnj}
L.~Susskind, ``{De Sitter Space has no Chords. Almost Everything is
  Confined.},'' \href{https://doi.org/10.22128/jhap.2023.661.1043}{{ JHAP}
  {\bfseries 3} no.~1, (2023) 1--30},
  \href{https://arxiv.org/abs/2303.00792}{{\ttfamily arXiv:2303.00792
  [hep-th]}}.

\bibitem{Rahman:2023pgt}
A.~A. Rahman and L.~Susskind, ``{Comments on a Paper by Narovlansky and
  Verlinde},'' \href{https://arxiv.org/abs/2312.04097}{{\ttfamily
  arXiv:2312.04097 [hep-th]}}.

\bibitem{Rahman:2024vyg}
A.~A. Rahman and L.~Susskind, ``{Infinite Temperature is Not So Infinite: The
  Many Temperatures of de Sitter Space},''
  \href{https://arxiv.org/abs/2401.08555}{{\ttfamily arXiv:2401.08555
  [hep-th]}}.

\bibitem{Rahman:2024iiu}
A.~A. Rahman and L.~Susskind, ``{$p$-Chords, Wee-Chords, and de Sitter
  Space},'' \href{https://arxiv.org/abs/2407.12988}{{\ttfamily arXiv:2407.12988
  [hep-th]}}.

\bibitem{Sekino:2025bsc}
Y.~Sekino and L.~Susskind, ``{Double-Scaled SYK, QCD, and the Flat Space Limit
  of de Sitter Space},'' \href{https://arxiv.org/abs/2501.09423}{{\ttfamily
  arXiv:2501.09423 [hep-th]}}.

\bibitem{Blommaert:2024ydx}
A.~Blommaert, T.~G. Mertens, and J.~Papalini, ``{The dilaton gravity hologram
  of double-scaled SYK},'' \href{https://arxiv.org/abs/2404.03535}{{\ttfamily
  arXiv:2404.03535 [hep-th]}}.

\bibitem{Blommaert:2024whf}
A.~Blommaert, A.~Levine, T.~G. Mertens, J.~Papalini, and K.~Parmentier, ``{An
  entropic puzzle in periodic dilaton gravity and DSSYK},''
  \href{https://arxiv.org/abs/2411.16922}{{\ttfamily arXiv:2411.16922
  [hep-th]}}.

\bibitem{Bossi:2024ffa}
L.~Bossi, L.~Griguolo, J.~Papalini, L.~Russo, and D.~Seminara, ``{Sine-dilaton
  gravity vs double-scaled SYK: exploring one-loop quantum corrections},''
  \href{https://arxiv.org/abs/2411.15957}{{\ttfamily arXiv:2411.15957
  [hep-th]}}.

\bibitem{Collier:2025pbm}
S.~Collier, L.~Eberhardt, and B.~M\"uhlmann, ``{The complex Liouville string:
  the gravitational path integral},''
  \href{https://arxiv.org/abs/2501.10265}{{\ttfamily arXiv:2501.10265
  [hep-th]}}.

\bibitem{Collier:2024kmo}
S.~Collier, L.~Eberhardt, B.~M\"uhlmann, and V.~A. Rodriguez, ``{The complex
  Liouville string},'' \href{https://arxiv.org/abs/2409.17246}{{\ttfamily
  arXiv:2409.17246 [hep-th]}}.

\bibitem{Sachdev1993}
S.~Sachdev and J.~Ye, ``Gapless spin-fluid ground state in a random quantum
  heisenberg magnet,'' \href{https://doi.org/10.1103/physrevlett.70.3339}{{
  Phys. Rev. Lett.} {\bfseries 70} no.~21, (1993) 3339--3342},
  \href{https://arxiv.org/abs/cond-mat/9212030}{{\ttfamily
  arXiv:cond-mat/9212030}}.

\bibitem{Kitaev1}
A.~Kitaev, ``A simple model of quantum holography (part 1),''.
  \url{https://online.kitp.ucsb.edu/online/entangled15/kitaev/}.

\bibitem{Kitaev2}
A.~Kitaev, ``A simple model of quantum holography (part 2),''.
  \url{https://online.kitp.ucsb.edu/online/entangled15/kitaev2/}.

\bibitem{Maldacena:2016hyu}
J.~Maldacena and D.~Stanford, ``{Remarks on the Sachdev-Ye-Kitaev model},''
  \href{https://doi.org/10.1103/PhysRevD.94.106002}{{ Phys. Rev. D} {\bfseries
  94} no.~10, (2016) 106002},
  \href{https://arxiv.org/abs/1604.07818}{{\ttfamily arXiv:1604.07818
  [hep-th]}}.

\bibitem{Goel:2023svz}
A.~Goel, V.~Narovlansky, and H.~Verlinde, ``{Semiclassical geometry in
  double-scaled SYK},'' \href{https://doi.org/10.1007/JHEP11(2023)093}{{ JHEP}
  {\bfseries 11} (2023) 093},
  \href{https://arxiv.org/abs/2301.05732}{{\ttfamily arXiv:2301.05732
  [hep-th]}}.

\bibitem{Stanford:2017thb}
D.~Stanford and E.~Witten, ``{Fermionic Localization of the Schwarzian
  Theory},'' \href{https://doi.org/10.1007/JHEP10(2017)008}{{ JHEP} {\bfseries
  10} (2017) 008}, \href{https://arxiv.org/abs/1703.04612}{{\ttfamily
  arXiv:1703.04612 [hep-th]}}.

\bibitem{Saad:2019lba}
P.~Saad, S.~H. Shenker, and D.~Stanford, ``{JT gravity as a matrix integral},''
  \href{https://arxiv.org/abs/1903.11115}{{\ttfamily arXiv:1903.11115
  [hep-th]}}.

\bibitem{Erickson:2000af}
J.~K. Erickson, G.~W. Semenoff, and K.~Zarembo, ``{Wilson loops in N=4
  supersymmetric Yang-Mills theory},''
  \href{https://doi.org/10.1016/S0550-3213(00)00300-X}{{ Nucl. Phys. B}
  {\bfseries 582} (2000) 155--175},
  \href{https://arxiv.org/abs/hep-th/0003055}{{\ttfamily
  arXiv:hep-th/0003055}}.

\bibitem{Drukker:2006ga}
N.~Drukker, ``{1/4 BPS circular loops, unstable world-sheet instantons and the
  matrix model},'' \href{https://doi.org/10.1088/1126-6708/2006/09/004}{{ JHEP}
  {\bfseries 09} (2006) 004},
  \href{https://arxiv.org/abs/hep-th/0605151}{{\ttfamily
  arXiv:hep-th/0605151}}.

\bibitem{Berkooz:2024ifu}
M.~Berkooz, R.~Frumkin, O.~Mamroud, and J.~Seitz, ``{Twisted times, the
  Schwarzian and its deformations in DSSYK},''
  \href{https://arxiv.org/abs/2412.14238}{{\ttfamily arXiv:2412.14238
  [hep-th]}}.

\bibitem{Okuyama:2023iwu}
K.~Okuyama, ``{High temperature expansion of double scaled SYK},''
  \href{https://doi.org/10.1016/j.physletb.2023.138036}{{ Phys. Lett. B}
  {\bfseries 843} (2023) 138036},
  \href{https://arxiv.org/abs/2304.01522}{{\ttfamily arXiv:2304.01522
  [hep-th]}}.

\bibitem{Blommaert:2025avl}
A.~Blommaert, A.~Levine, T.~G. Mertens, J.~Papalini, and K.~Parmentier,
  ``{Wormholes, branes and finite matrices in sine dilaton gravity},''
  \href{https://arxiv.org/abs/2501.17091}{{\ttfamily arXiv:2501.17091
  [hep-th]}}.

\end{thebibliography}\endgroup
\bibliographystyle{utphys}

\end{document}